\documentclass{nature}

\usepackage{bm}
\usepackage{graphicx}
\usepackage{color}
\usepackage[normalem]{ulem}

\newcommand{\apj}{Astrophys.~J.}
\newcommand{\apjl}{Astrophys.~J.}
\newcommand{\mnras}{Mon.~Not.~R.~Astron.~Soc.}
\newcommand{\aap}{Astron.~Astrophys.}

\newcommand{\nat}{Nature}

\DeclareMathSymbol{\varOmega}{\mathord}{letters}{"0A}
\DeclareMathSymbol{\varSigma}{\mathord}{letters}{"06}

\title{Rapid planetesimal formation in turbulent circumstellar discs}

\author{Anders Johansen$^1$, Jeffrey S. Oishi$^{2,3}$, Mordecai-Mark Mac
Low$^{2,1}$, Hubert Klahr$^1$, Thomas Henning$^1$ \& Andrew Youdin$^4$}

\begin{document}

\maketitle

\begin{affiliations}
  \item Max-Planck-Institut f\"ur Astronomie, K\"onigstuhl 17, D-69117
      Heidelberg, Germany
  \item Department of Astrophysics, American Museum of Natural
      History, 79th Street at Central Park West, New
      York, NY 10024-5192, USA
  \item also Department of Astronomy, University of Virginia,
  Charlottesville, VA, USA
  \item Canadian Institute for Theoretical Astrophysics, University of Toronto, 60 St. George Street, Toronto, Ontario M5S 3H8, Canada
\end{affiliations}

\begin{abstract}

\noindent{\bf
  The initial stages of planet formation in circumstellar gas discs proceed via
  dust grains that collide and build up larger and larger
  bodies\cite{Safronov1969}. How this process continues from metre-sized
  boulders to kilometre-scale planetesimals is a major unsolved
  problem\cite{Dominik+etal2007}: boulders stick together
  poorly\cite{Benz2000}, and spiral into the protostar in a few hundred orbits
  due to a head wind from the slower rotating gas\cite{Weidenschilling1977}.
  Gravitational collapse of the solid component has been suggested to overcome
  this barrier\cite{Safronov1969,GoldreichWard1973,YoudinShu2002}. Even low
  levels of turbulence, however, inhibit sedimentation of solids to a
  sufficiently dense midplane
  layer\cite{WeidenschillingCuzzi1993,Dominik+etal2007}, but turbulence must be
  present to explain observed gas accretion in protostellar
  discs\cite{Hartmann1998}. Here we report the discovery of efficient
  gravitational collapse of boulders in locally overdense regions in the
  midplane. The boulders concentrate initially in transient high pressures in
  the turbulent gas\cite{JohansenKlahrHenning2006}, and these concentrations
  are augmented a further order of magnitude by a streaming
  instability\cite{YoudinGoodman2005,JohansenHenningKlahr2006,
  JohansenYoudin2007} driven by the relative flow of gas and solids. We find
  that gravitationally bound clusters form with masses comparable to dwarf
  planets and containing a distribution of boulder sizes.  Gravitational
  collapse happens much faster than radial drift, offering a possible path to
  planetesimal formation in accreting circumstellar discs.
}

\end{abstract}

Planet formation models typically treat turbulence as a diffusive process that
opposes the gravitational sedimentation of solids to a high density midplane
layer in circumstellar discs\cite{WeidenschillingCuzzi1993,Cuzzi+etal1993}.
Recent models of solids moving in turbulent gas reveal that the turbulent
motions not only mix them, but also concentrate metre-sized boulders in the
transient gas overdensities\cite{JohansenKlahrHenning2006} formed in
magnetorotational turbulence\cite{BalbusHawley1998}, in giant gaseous
vortices\cite{BargeSommeria1995,FromangNelson2005}, and in spiral arms of
self-gravitating discs\cite{Rice+etal2006}. Short-lived eddies at the
dissipation scale of forced turbulence concentrate smaller millimetre-sized
solids\cite{Cuzzi+etal2001}.

Some simulations mentioned
above\cite{JohansenKlahrHenning2006,JohansenHenningKlahr2006,JohansenYoudin2007}
were performed with the Pencil Code, which solves the magnetohydrodynamic (MHD)
equations on a three-dimensional grid for a gas that interacts through drag
forces with boulders. Boulders are represented as superparticles with
independent positions and velocities, each having the mass of a huge number of
boulders but the aerodynamic behaviour of a single boulder.  We have now
further developed the Pencil Code to include a fully parallel solver for the
gravitational potential of the particles (see Supplementary Information). The
particle density is mapped on the grid using the Triangular Shaped Cloud
assignment scheme\cite{HockneyEastwood1981} and the gravitational potential of
the solids is found using a Fast Fourier Transform method\cite{Gammie2001}.
This allows us, for the first time, to simulate the dynamics of
self-gravitating solid particles in magnetised, three-dimensional turbulence.

We model a corotating, local box with linearised Keplerian shear that straddles
the protoplanetary disc midplane and orbits the young star at a fixed
distance.  Periodic boundary conditions are applied. An isothermal equation of
state is used for the gas, while the induction equation is solved under the
ideal MHD assumption of high conductivity. Magnetorotational
instability\cite{BalbusHawley1998} drives turbulence in Keplerian discs with
sufficient ionisation\cite{Gammie1996}, producing in our unstratified models
turbulence with Mach number ${\rm Ma}\approx0.05$ and viscosity
$\alpha\approx10^{-3}$, a realistic value to explain observed accretion
rates\cite{Hartmann1998}. The ionisation fraction in the dense midplanes of
protoplanetary discs may be insufficient for the gas to couple with the
magnetic field to drive magnetorotational instability\cite{Gammie 1996}. In the
Supplementary Information we therefore describe unmagnetised models as well.

Solid objects orbit the protostar with Keplerian velocity $v_{\rm K}$ in the
absence of gas drag.  A radial pressure gradient partly supports the gas,
however, so it orbits at sub-Keplerian velocity, with $\Delta v\equiv v_{\rm
g}-v_{\rm K}<0$. As a result, large (approximately metre-sized) solid objects
feel a strong head wind that causes them to drift radially
inwards\cite{Weidenschilling1977} with a maximum drift velocity $\Delta v$.
They also feel gas drag as they fall toward the disc midplane in the effective
gravity field of the star.  A sedimentary midplane layer forms with a width
determined by a balance between settling and turbulent
diffusion\cite{WeidenschillingCuzzi1993,Cuzzi+etal1993}.

We present three types of models: (1) without self-gravity, with $128^3$ zones
and $2\times10^6$ particles, run for 100 orbits, to study the interplay between
the streaming instability\cite{YoudinGoodman2005} and concentration by
transient high pressures; (2) with self-gravity and boulder collisions, with
$256^3$ zones and $8\times10^6$ particles run for 27 orbits, to study
gravitational collapse; and (3) models with self-gravity but no
magnetorotational turbulence (presented in the Supplementary Information).
Magnetorotational turbulence is given 10 orbits to reach steady state before we
turn on drag force and vertical gravity, to avoid any influence of the initial
conditions on the sedimented midplane layer.  We fix the global solids-to-gas
bulk density ratio at the canonical galactic value of $\epsilon_0=0.01$, but
two values of the radial drift are considered: low drift with $\Delta
v=-0.02c_{\rm s}$, where $c_{\rm s}$ is the isothermal sound speed, and
moderate drift with $\Delta v=-0.05c_{\rm s}$, depending on the assumed radial
pressure support (values up to $\Delta v=-[0.2\ldots0.5]c_{\rm s}$ are
possible\cite{Cuzzi+etal1993}, but are not considered here).

\begin{figure}
  \caption{
    Topography of the sedimented particle layer in models without self-gravity
    or collisional cooling. a) The azimuthally averaged vertical column density
    $\varSigma_{\rm p}$ of metre-sized boulders (with $\varOmega_{\rm K}
    \tau_{\rm f}=1$) as a function of radial coordinate $x$ and time $t$, in a
    model where the particles feel gas drag, but the gas does not feel drag
    from the particles. Radial drift is evident from the tilted bands
    (particles crossing the inner boundary reappear at the outer). Transient
    regions of mildly increased gas pressure temporarily concentrate boulders.
    The gas orbits slightly slower on the outer edge of these high pressure
    regions and slightly faster on the inner edge, resulting in a differential
    head wind that forces boulders towards their
    centres\cite{HaghighipourBoss2003,JohansenKlahrHenning2006}. b) Including
    the drag force from the particles on the gas allows for the development of
    the streaming instability, seeded by the existing radial density
    enhancements. The streaming instability occurs where the collective drag
    force of the solids forces the gas to locally move with an orbital speed
    that is closer to Keplerian, reducing the gaseous head wind that otherwise
    causes boulders to drift radially. Solids then drift into already overdense
    regions from further out, causing runaway growth in the local bulk density
    of solids. c) The column density when the radial pressure support is
    increased from $\Delta v=-0.02c_{\rm s}$ to $-0.05c_{\rm s}$. Radial
    density enhancements become narrower and shorter-lived due to downstream
    erosion of the overdensities by the stronger radial drift. d) The maximum
    solid-to-gas ratio on the grid as a function of time. The average
    solids-to-gas ratio in the midplane is $0.5$, whereas the maximum reaches
    well over ten times higher values in transient high pressure regions
    (yellow) and several hundred times higher values when the streaming
    instability is active (orange and blue).
    \\
    \begin{center}
      \includegraphics[width=13.6cm]{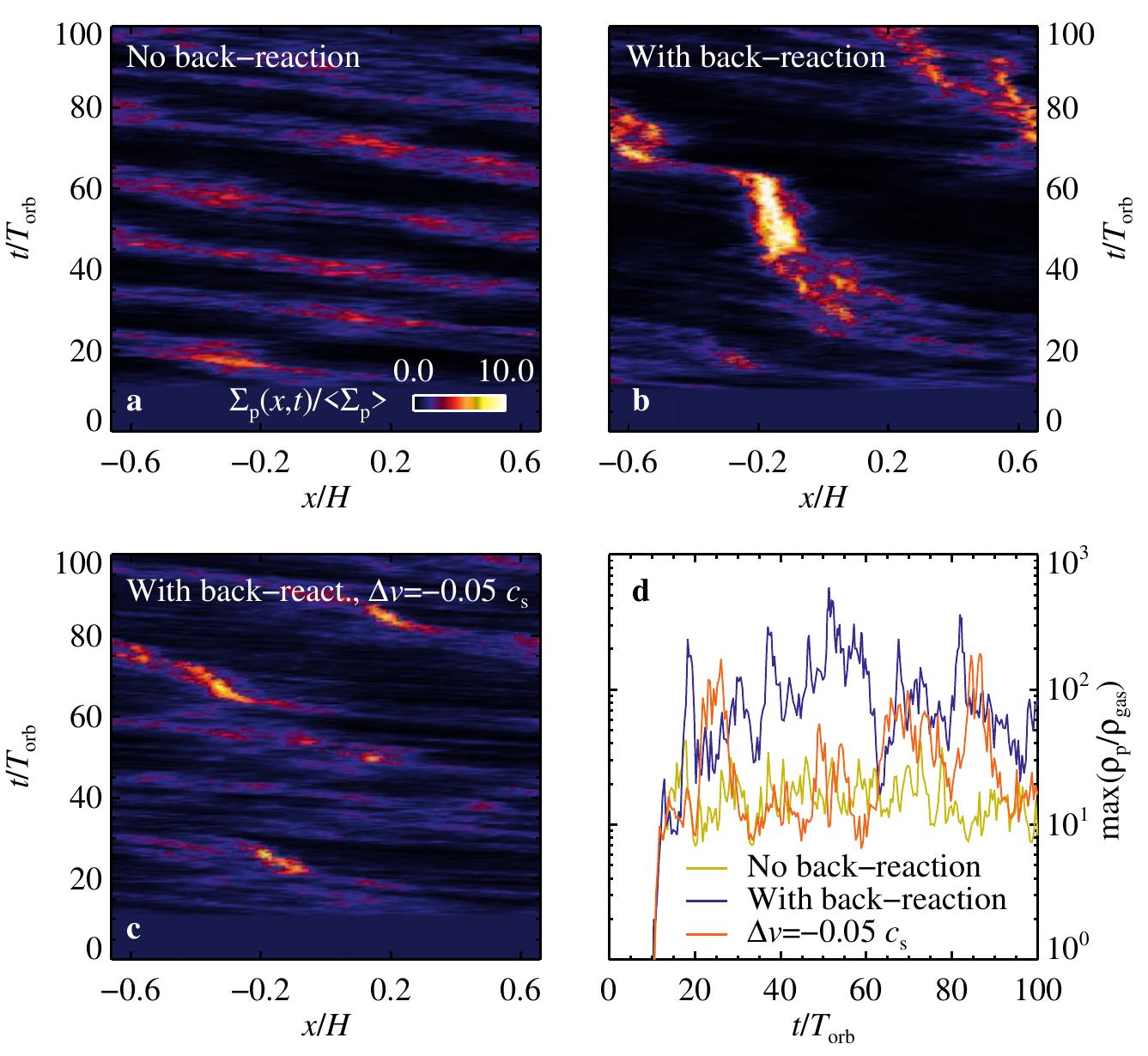}
    \end{center}
    (this is Figure 1)\newpage
  \label{f:rhop_overview}
} 
\end{figure}

For the simulations without self-gravity we consider a fixed particle size
parameterised by the dimensionless friction time $\varOmega_{\rm K}\tau_{\rm
f}=1.0$, where $\varOmega_{\rm K}$ is the local Keplerian rotation frequency
and $\tau_{\rm f}$ is the time-scale over which gas and solids reach equal
velocity. At an orbital distance $r=5\,{\rm AU}$ this corresponds to boulders
of approximately one metre in diameter. Figure~\ref{f:rhop_overview} shows the
space-time topography of the sedimented midplane layer. The streaming
instability increases the density of boulders in regions where they have
already been concentrated by transient high
pressures\cite{JohansenKlahrHenning2006}. Increasing radial pressure support
from $\Delta v=-0.02c_{\rm s}$ to $-0.05c_{\rm s}$ reduces the concentration by
streaming instability, although the local solids-to-gas density ratio still
reaches 200.

Gravitational collapse of discrete solid objects produces virialised clusters
unable to contract further\cite{Tanga+etal2004} in the absence of mechanisms to
dynamically cool the cluster--that is, reduce the local rms speed. Two
processes that we consider can be important: drag force cooling and collisional
cooling. Drag force cooling occurs because part of the kinetic energy exchanged
between the particles and the gas is dissipated. Collisional cooling is
produced by the highly inelastic collisions between boulders, transferring
kinetic energy to heat and deformation. Collisional cooling occurs generally in
simulations of resolved collisions in planetary rings\cite{Salo1992}. In the
Supplementary Information we describe how we treat collisional cooling
numerically in the self-gravitating simulations by damping the rms speed of the
particles in each grid cell on a collisional time-scale. We have found that in
the absence of collisional cooling, gravitational collapse still proceeds if
the total surface density (of solids and gas) is augmented by 50\%.
Collisional cooling is thus not a prerequisite of the collapse, but does allow
it to occur in somewhat less massive discs. We ignore all other effects of the
collisions, such as coagulation and collisional fragmentation. Collisional
cooling and self-gravity are turned on after 20 orbits in the self-gravitating
simulations.

Our chosen scale-height-to-radius ratio of $H/r=0.04$ gives a gas temperature
of $T=80\,{\rm K}$ at an orbital radius of $r=5\,{\rm AU}$.  We choose for the
$256^3$ self-gravitating run the uniform gas volume density to be
consistent with the midplane of a disc with surface density of $\varSigma_{\rm
gas}=300$~g~cm$^{-2}$. This corresponds to approximately twice the minimum mass
solar nebula (MMSN) at 5 AU from the (proto-)Sun. An alternative theory for
giant planet formation, the disc instability
hypothesis\cite{Boss1997,Mayer+etal2002}, requires column densities at least 20
times higher than the MMSN for gravitational fragmentation of the gaseous
component of the disc to occur.

We have examined the numerical convergence of our models with resolutions
ranging from $64^3$ to $256^3$ zones (see Supplementary Information).  The peak
particle density on the grid increases with increasing resolution, because of
less smoothing in the particle-mesh scheme at higher resolution, resulting in a
decrease in the column density threshold for gravitational collapse. Although
we have not yet fully converged, our results appear to provide good {\em upper
limits} to the column density for which collapse can occur.  For the
self-gravitating simulation we consider boulders with friction times
distributed among $\varOmega_{\rm K}\tau_{\rm f}=0.25,0.50,0.75,1.00$. At
$r=5\,{\rm AU}$ in our chosen disc model, these correspond to radii of
15--60~cm. Consideration of multiple boulder sizes is vital since differential
aerodynamic behaviour could inhibit gravitational
instabilities\cite{Weidenschilling1995}. The size range covers roughly half of
the two orders of magnitude in particle radius produced by coagulation of
microscopic grains\cite{DullemondDominik2005}. Smaller particles are ignored
since they are unlikely to separate from the gas and participate in
gravitational collapse. In case of widespread collisional fragmentation, e.g.\
in the warmer terrestrial planet formation region, up to 80\% of the solid
material may be bound in small
fragments\cite{Weidenschilling1997,DullemondDominik2005}, in which case we must
implicitly assume an augmentation in solids-to-gas ratio of up to $5$.

In our self-gravitating model we set $\Delta v=-0.02c_{\rm s}$, but show in the
Supplementary Information that gravitationally bound clusters also form for
$\Delta v=-0.05c_{\rm s}$, with a factor of two increase in column density
threshold. The Supplementary Information also documents that typical boulder
collisions happen at speeds below the expected destruction
threshold\cite{Benz2000}. We caution, however, that material properties, and
thus destruction thresholds, of the boulders are poorly known. Higher
resolution studies, and an improved analytical theory of collision speeds that
takes into account epicyclic motion, will be needed to determine whether
collision speeds have converged, given an unexplained factor 3 difference for
$\varOmega_{\rm K}\tau_{\rm f}=1$ particles between typical relative speeds
within cells ($\approx5$~m~s$^{-1}$) and the expected collision speed of
well-mixed particles.

\begin{figure}
  \caption{
    Time series of the collapse of overdense seeds into gravitationally bound
    boulder clusters.  The central panel shows the column densities of the four
    different sizes of boulders (in units of the mean column density of each
    size) plotted independently at a time just before self-gravity is turned
    on. All four particle sizes have concentrated at similar locations, an
    important prerequisite for the subsequent gravitational collapse.  The
    surrounding panels show a time series of total column density of solids, in
    the radial-azimuthal ($x$-$y$) plane of the disc, summed over all particle
    sizes, starting from the upper left and progressing clockwise. Values are
    normalised to the average value across the grid (see colour bars in upper
    right panel). Times are given in orbital times $T_{\rm orb}$ after
    self-gravity is turned on.  Inset in each panel is an enlargement of a
    square region (indicated in the main panel) centred around the Hill sphere
    of the most massive cluster in the simulation, represented by the white
    circle.  These inserts show the log of the column density ratio (see colour
    bar in upper right panel) to capture the extreme values reached. Overdense
    bands initially contract radially, forming thin filaments with densities
    high enough for a full non-axisymmetric collapse into gravitationally bound
    clumps to take place. As time progresses, the Hill sphere increases in
    radius as the clusters grow in mass by accreting boulders from the
    turbulent flow (see Supplementary Video for an animation of this
    simulation).
    \\
    \begin{center}
      \includegraphics[width=\linewidth]{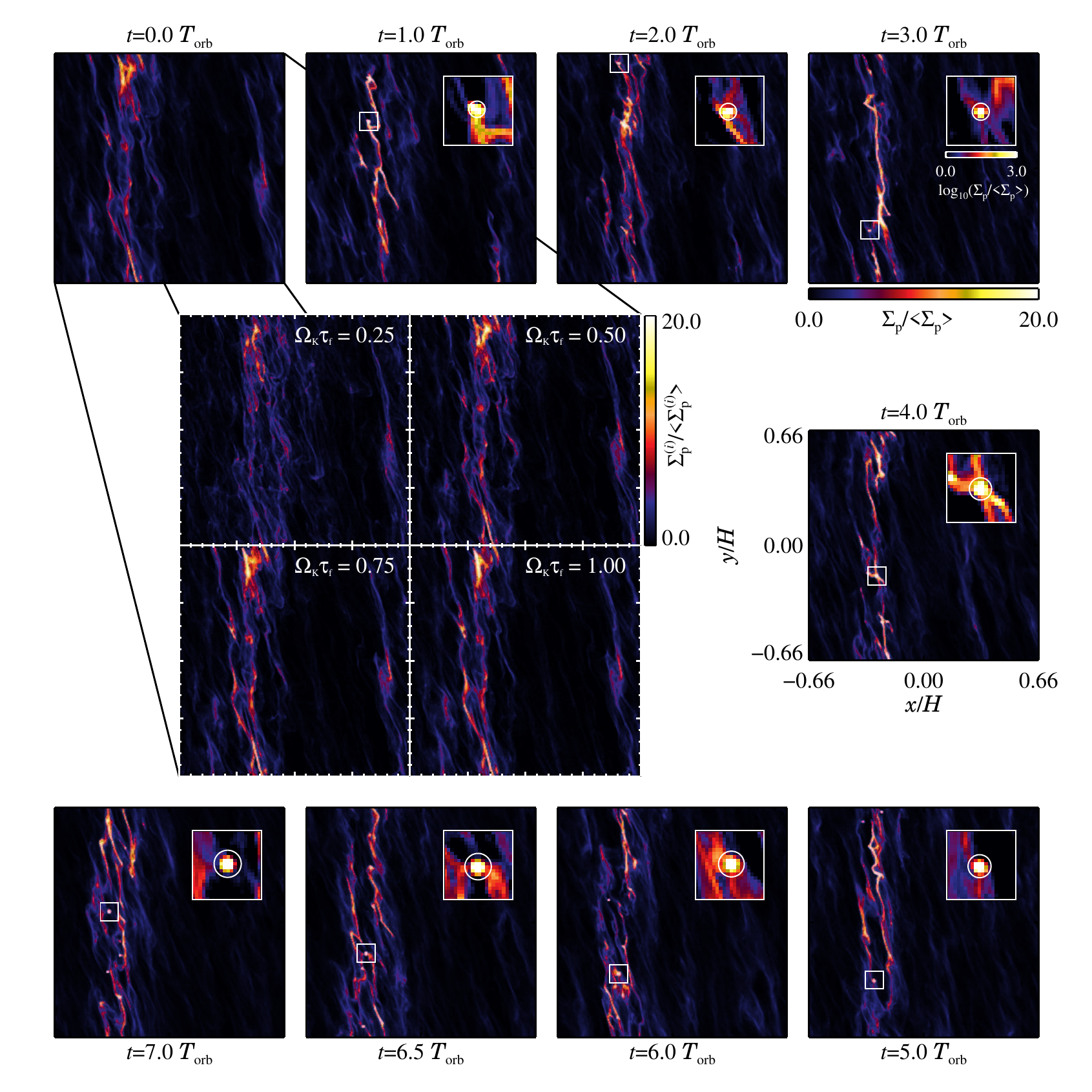}
    \end{center}
    (this is Figure 2)\newpage
    \label{f:sg_time_ser_multi_wrap}
} 
\end{figure}

The development of gravitational instability in the $256^3$ run is shown in
Figure~\ref{f:sg_time_ser_multi_wrap}.  The four different boulder sizes have
accumulated in the same regions ({\em central panel}) already before
self-gravity is turned on, demonstrating that differential drift does not
prevent density enhancement by streaming instability.  Gravitationally bound
boulder clusters form, with a Hill radius---within which the gravity of the
cluster dominates over tidal forces from the central star---that increases
steadily with time ({\em inserts}) as the clusters accrete boulders from the
surrounding flow.

\begin{figure}
  \caption{
    Mass accretion onto a gravitationally bound cluster. The plot shows the
    maximum bulk density of solids as a function of time, normalised by the
    average gas density. Drag force and vertical gravity are turned on at
    $t=-10$, while self-gravity and collisional cooling are turned on at $t=0$.
    The density increases monotonically after the onset of self-gravity because
    gravitationally bound clusters of boulders form in the mid-plane.  After
    only seven orbits peak densities in these clusters approach $10^4\rho_{\rm
    g}$ or a million times the average boulder density in the disc.  The
    coloured bars show the mass contained within the most massive Hill sphere
    in the box, in units of the mass of the 970 km radius dwarf planet Ceres
    ($M_{\rm Ceres}=9.5\times10^{23}\,{\rm g}$). The most massive cluster
    accretes about $0.5 M_{\rm Ceres}$ per orbit (the entire box contains a
    total boulder mass of $50 M_{\rm Ceres}$).  The cluster consists of
    approximately equal fractions of the three larger boulder sizes. The
    smallest size, with $\varOmega_{\rm K}\tau_{\rm f}=0.25$, is initially
    underrepresented with a fraction of only $15\%$ because of the stronger
    aerodynamic coupling of those particles to the gas, but the fraction of
    small particles increases with time as the cluster grows massive enough to
    attract smaller particles as well.  The mean free path inside the bound
    clusters is shorter than the size of the cluster, so any fragments formed
    in catastrophic collisions between the boulders will be swept up by the
    remaining boulders before being able to escape the cluster (see
    Supplementary Information).
    \\
    \begin{center}
      \includegraphics{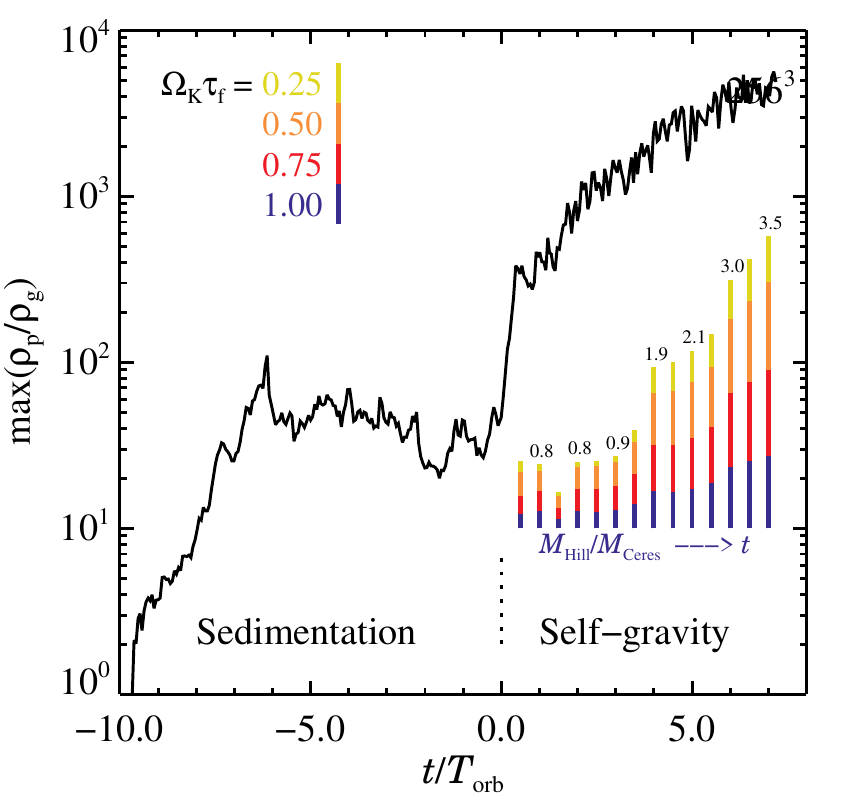}
    \end{center}
    (this is Figure 3)
\label{f:rhopmax_t_sg}
} 
\end{figure}

We show in Figure~\ref{f:rhopmax_t_sg} the peak density and the mass of the
most massive gravitationally bound cluster as a function of time. The cluster
consists of particles of all four sizes, demonstrating that different boulder
sizes can indeed take part in the same gravitational collapse, despite their
different aerodynamical properties and drift behaviour. At the end of the
simulation the most massive cluster contains 3.5 times the mass of the dwarf
planet Ceres. The cluster mass agrees roughly with standard estimates from
linear gravitational instability\cite{GoldreichWard1973}, when applied at
$r=5\,{\rm AU}$ to the locally enhanced column densities (see Supplementary
Information).  

Models lacking magnetic fields, and thus magnetorotational turbulence, are
described in the Supplementary Information. Here sedimentation occurs
unhindered until the onset of Kelvin-Helmholtz instabilities driven by the
vertical shear of gas velocity above the midplane boulder layer. Strong boulder
density enhancements in the mid-plane layer nevertheless still form, although
for moderate drift an increase of the solids-to-gas ratio from $0.01$ to $0.03$
(perhaps possible due to radial variation in boulder drift
speeds\cite{YoudinShu2002} or photoevaporation of the
gas\cite{ThroopBally2005}) was needed to obtain strong clumping.
Magnetorotational turbulence thus has a positive effect on the mid-plane
layer's ability to gravitationally collapse, although collapse can occur
without it as well.

The Supplementary Information also includes a model with an
adiabatic equation of state and explicit gas heating due to energy
dissipated by drag and inelastic collisions. We find that gas heating
does not prevent collapse. The maximum temperature reached is not even high
enough to melt ice, although that may change with the formation of massive
bodies with escape velocity near the sound speed.

Our proposed path to planetesimal formation depends crucially on the existence
of a dense sedimentary layer of boulders. Future investigations should focus on
the formation and survival of such layers in light of processes like
coagulation, collisional fragmentation and erosion\cite{Weidenschilling1997}.
Especially important are higher resolution studies of collision speeds and an
improved analytical theory of collisions that includes the epicyclic motion of
particles.

\begin{addendum}
 \item [Supplementary Information] is linked to the online version of the paper
   at www.nature.com/nature.
 \item This collaboration was made possible through the support of the
   Annette Kade Graduate Student Fellowship Program at the American
   Museum of Natural History. J.S.O. was supported by the US National
   Science Foundation, as was M.-M.M.L. in part. We thank J. Cuzzi for an
   e-mail discussion about the role of cooling on the gravitational collapse.
 \item[Competing Interests] The authors declare that they have no competing
   financial interests.
 \item[Correspondence] Correspondence and requests for materials should be
   addressed to A.J.~(email: johansen@mpia.de).
\end{addendum}

\end{document}